\documentclass[preprint,showpacs,amsmath,amssymb]{revtex4}
\usepackage{amssymb}

\usepackage{graphicx}
\usepackage{subfig}
\usepackage{tabularx}
\usepackage{dcolumn}
\usepackage{bm}

\begin{document}

\title{Evidence for Some New Hyperon Resonances
- to be Checked by $K_L$ Beam Experiments}

\author{B.~S.~Zou}

\affiliation{State Key Laboratory of Theoretical Physics, Institute
of Theoretical Physics, Chinese Academy of Sciences, Beijing 100190,
China}

\begin{abstract}

Quenched and unquenched quark models predict very different patterns
for the spectrum of the low excited hyperon states. Evidence is
accumulating for the existence of some new hyperon resonances, such
as a $\Sigma^*$ of spin-parity $J^P=1/2^-$ around 1400 MeV instead
of 1620 MeV as listed in PDG, a new $\Sigma(1540)3/2^-$ resonance, a
new narrow $\Lambda(1670)3/2^-$ resonance and a new
$\Lambda(1680)3/2^+$ resonance. All these new hyperon resonances fit
in the predicted pattern of the unquenched quark models very well.
It is extremely important to check and establish the spectrum of
these low excited hyperon states by the proposed $K_L$ beam
experiments at JLAB.

\end{abstract}
 \maketitle{}

\section{Why hyperon resonances ?}

Creation of quark-anti-quark pairs from gluon field plays a crucial
role for understanding quark confinement and hadron spectroscopy. In
the classical quenched quark model for a $q_1\bar q_1$ meson, the
$q_1$ quark cannot be separated from the $\bar q_1$ anti-quark due
to a infinitely large confinement potential. But in realty, we know
the $q_1$ and $\bar q_1$ can be easily separated from each other by
creation of another quark-anti-quark pair $q_2\bar q_2$ to decay to
two mesons,  $q_1\bar q_2$ and $q_2\bar q_1$. With the creation of
the $q_2\bar q_2$, instead of forming two colorless mesons, the
system could also exist in the form of a tetra-quark state
$[q_1q_2][\bar q_1\bar q_2]$. Therefore both lattice QCD and quark
models should go beyond the quenched approximation which ignore the
creation of quark-anti-quark pairs.

\begin{figure}[htbp]
 \includegraphics*[width=14cm]{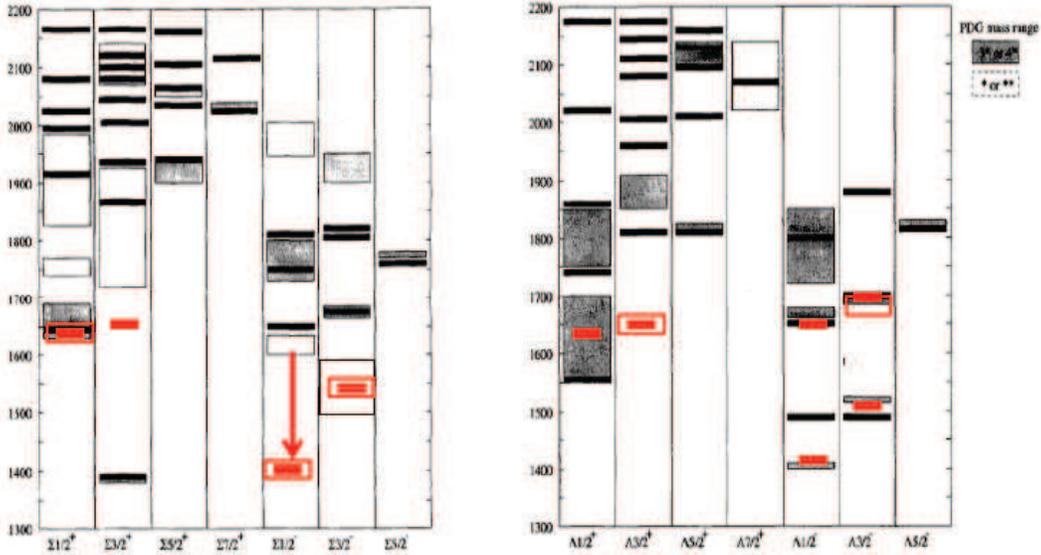}
\caption{Prediction of the lowest penta-quark states with
$J^P=1/2^\pm,3/2^\pm$~\cite{pentq1,pentq2} (red solid) compared with
those from the classical quenched $qqq$ model~\cite{Capstick0}
(black solid). The black boxes are experimental results from PDG
while the red box are from recent new analyses. }\label{fig:graph}
\end{figure}

Quenched $qqq$ quark models and unquenched $qqq\leftrightarrow
qqqq\bar q$ quark models give very different predictions for the
hyperon spectroscopy. For example, for the $J^P={1\over 2}^-$ SU(3)
nonet partners of the $N(1535)$ and $\Lambda(1405)$. While quenched
quark models~\cite{Capstick0,capstick,gloz,loring} predict the
$J^P={1\over 2}^-$ $\Sigma$ and $\Xi$ resonances to be around 1650
MeV and 1760 MeV, respectively, the unquenched quark
models~\cite{pentq1,pentq2,zoupent} expect them to be around 1400
MeV and 1550 MeV, respectively, a meson-soliton bound-state approach
of the Skyrme model~\cite{Oh} and other meson-baryon dynamical
models~\cite{Kanchan11,Ramos} predict them to be around 1450 MeV and
1620 MeV, respectively. In Fig.\ref{fig:graph}, we show prediction
of the lowest penta-quark states with
$J^P=1/2^\pm,3/2^\pm$~\cite{pentq1,pentq2} (red solid) compared with
those from the classical quenched $qqq$ model~\cite{Capstick0}
(black solid). The major differences are that the lowest penta-quark
hyperon states with $J^P=1/2^-$ and $3/2^+$ are about 200 MeV lower
those from the classical quenched $qqq$ models~\cite{Capstick0}.

Although various phenomenological models give distinguishable
predictions for the lowest excited hyperon states, most of them are
not experimentally established or even listed in PDG~\cite{pdg}.
Most of our knowledge for the hyperon resonances came from analyses
of old ${\overline K}N$ experiments in the 1970s~\cite{pdg}. In the
new century,  some new measurements from Crystal Ball
(CB)~\cite{CBdata0,CBdata1,CBdata2}, LEPS~\cite{LEPS} and
CLAS~\cite{CLAS} have started to provide us new information on
$\Sigma^*$ and $\Lambda^*$ resonances. It is crucial to use them to
clarify the spectrum of low-lying hyperon resonances to pin down the
underlying dynamics for baryon spectrum and structure. Recent
analyses of these new data together with old data reveal some
interesting new features of the low-lying excited hyperon states.
Here I will give a brief review of these new results and discuss
about their further confirmation from the proposed $K_L$ beam and
other experiments.

\section{New results on $\Sigma^*$ and $\Lambda^*$ resonances}

\subsection{On the lowest $\Sigma^*$ resonances with negative parity}

The lowest $\Sigma^*$ resonances with $J^P=1/2^-$ or $3/2^-$ are
still far from established. There is a $\Sigma(1620){1\over 2}^-$
listed as a 2-star resonance in the previous versions of PDG tables
and downgraded to 1-star in the newest version~\cite{pdg}. There is
also a $\Sigma(1580){3\over 2}^-$ listed as 1-star
resonance~\cite{pdg}.

The $\Sigma(1620){1\over 2}^-$ seems supporting the prediction of
quenched quark models. However, for the 2-star
$\Sigma(1620)\frac{1}{2}^-$ resonance, only four
references~\cite{16208,16207,16202,16201} are listed in PDG tables
with weak evidence for its existence. Among them, Ref.~\cite{16208}
and Ref.~\cite{16207} are based on multi-channel analysis of the
$\overline{K}N$ reactions.  Both claim evidence for a
$\Sigma(\frac{1}{2}^-)$ resonance with mass around 1620 MeV, but
give totally different branching ratios for this resonance.
Ref.~\cite{16208} claims that it couples only to $\pi\Lambda$ and
not to $\pi\Sigma$ while Ref.~\cite{16207} claims the opposite way.
Both analyses do not have $\Sigma(1660)\frac{1}{2}^+$ in their
solutions. However, Ref.~\cite{NPB94} shows no sign of
$\Sigma(\frac{1}{2}^-)$ resonance between 1600 and 1650 MeV through
analysis of the reaction $\overline{K}N\rightarrow\Lambda\pi$ with
the c.m. energy in the range of 1540-2150 MeV, instead it suggests
the existence of $\Sigma(1660)\frac{1}{2}^+$. Later multi-channel
analyses of the $\overline{K}N$ reactions support the existence of
the $\Sigma(1660)\frac{1}{2}^+$ instead of
$\Sigma(1620)\frac{1}{2}^-$~\cite{pdg}. In Ref.~\cite{16202}, the
total cross sections for $K^-p$ and $K^-n$ with all proper final
states are analyzed and indicate some $\Sigma$ resonances near 1600
MeV without clear quantum numbers. Ref.~\cite{16201} analyzes the
reaction $K^-n\rightarrow\pi^-\Lambda$ and gets two possible
solutions, with one solution indicating a $\Sigma({1\over 2}^-)$
near 1600 MeV, and the other showing no resonant structure below the
$\Sigma(1670)$. So all these claims of evidence for the
$\Sigma(1620){1\over 2}^-$ are very shaky. Instead, some re-analyses
of the $\pi\Lambda$ relevant data suggest that there may exist a
$\Sigma({1\over 2}^-)$ resonance around 1380 MeV~\cite{Wujj}, which
supports the prediction of unquenched quark
models~\cite{pentq1,pentq2}. This is supported by the new CLAS data
on $\gamma p\to K \Sigma\pi$~\cite{CLAS}, although a more delicate
analysis~\cite{Roca:2013cca} of the data suggests the resonant peak
to be at a higher mass around 1430 MeV.

For the study of $\Sigma$ resonances, the ${\bar K} N\to\pi\Lambda$
reaction is the best available channel, where the s-channel
intermediate states are purely hyperons with strangeness $S=-1$ and
isospin $I=1$. Recently, high statistic new data for the reaction
$K^-p\rightarrow\pi^0\Lambda$ are presented by the Crystal Ball
collaboration with the c.m. energy of 1560-1676 MeV for both
differential cross sections and $\Lambda$
polarizations~\cite{CBdata1}. In order to clarify the status of the
$\Sigma(1620)\frac{1}{2}^-$ and the $\Sigma(1660)\frac{1}{2}^+$, we
analyzed the differential cross sections and $\Lambda$ polarizations
for both $K^-p\to\pi^0\Lambda$ and $K^-n\to\pi^-\Lambda$ reactions
with an effective Lagrangian approach, using the new Crystal Ball
data on $K^-p\to\pi^0\Lambda$ with the c.m. energy of 1560-1676
MeV~\cite{CBdata1}, and the $K^-n\to\pi^-\Lambda$ data of
Ref.~\cite{16201} with the c.m. energy of 1550-1650~MeV, where the
evidence of the $\Sigma(1620)\frac{1}{2}^-$ was claimed. The new
Crystal Ball data clearly shows that the Crystal Ball $\Lambda$
polarization data demand the existence of a $\Sigma$ resonance with
$J^P=\frac{1}{2}^+$ and mass near 1635~MeV~\cite{pzgao}, compatible
with $\Sigma(1660)\frac{1}{2}^+$ listed in PDG, while the
$\Sigma(1620){1\over 2}^-$ is not needed by the data. The
differential cross sections alone cannot distinguish the two
solutions with either $\Sigma(1660)\frac{1}{2}^+$ or
$\Sigma(1620)\frac{1}{2}^-$.

This analysis also suggests a possible $\Sigma(\frac{3}{2}^-)$
resonance with mass around 1542 MeV and width about 25.6 MeV. This
seems consistent with the resonance structure $\Sigma(1560)$ or
$\Sigma(1580){3\over 2}^-$ in PDG and compatible with expectation
from penta-quark model~\cite{pentq1}. Ref.~\cite{D13} also proposes
a $\Sigma({3\over 2}^-)$ resonance with mass around 1570 MeV and
width about 60 MeV from $\overline{K}N\pi$ system.

After our analysis, there were three
groups~\cite{Zhang:2013sva,Kamano:2015hxa,Fernandez-Ramirez:2015tfa}
having made more sophisticated coupled channel analysis of the $\bar
KN$ scattering data including those from the Crystal Ball
experiment. The newest analysis~\cite{Fernandez-Ramirez:2015tfa}
gives roughly consistent results for the lowest $\Sigma^*(1/2^\pm)$
resonances as ours. In both analyses, there is no
$\Sigma(1620)1/2^-$. While in our analysis, the $\Sigma(1635)1/2^+$
is definitely needed, in Ref.\cite{Fernandez-Ramirez:2015tfa}, the
$\Sigma(1635)1/2^+$ is split to two $1/2^+$ resonances:
$\Sigma(1567)$ and $\Sigma(1708)$. The other two analyses claim the
need of the $\Sigma(1620)1/2^-$, but with much lower energy at 1501
MeV~\cite{Zhang:2013sva} and 1551 MeV~\cite{Kamano:2015hxa},
respectively.

For the lowest $\Sigma^*(3/2^-)$, Ref.\cite{Kamano:2015hxa} gives a
similar result as ours with mass around 1550 MeV.
Refs.\cite{Zhang:2013sva,Fernandez-Ramirez:2015tfa} give a higher
mass around 1670 MeV.

So there are strong evidences for the lowest $\Sigma^*(1/2^-)$ to be
in the range of $1380\sim 1500$ MeV and the lowest $\Sigma^*(3/2^-)$
to be around 1550 MeV. But this is not conclusive.

\subsection{On the lowest $\Lambda^*(3/2^\pm)$ resonances }

Many studies have been carried out to investigate the $\Lambda$
resonances. Oset et al.~\cite{Oset1,Oset2} used a chiral unitary
approach for the meson-baryon interactions and got two $J^P={1\over
2}^-$ resonances with one mass near 1390 MeV and the other around
1420 MeV. They believe the well established $\Lambda(1405){1\over
2}^-$ resonance listed in PDG~\cite{pdg} is actually a superposition
of these two ${1\over 2}^-$ resonances. Manley et
al.~\cite{Zhang:2013sva} and Kamano et al.~\cite{Kamano:2015hxa}
made multichannel partial-wave analysis of $\overline{K}N$ reactions
and got results with some significant differences. Zhong et
al.~\cite{xhZhong} analyzed the $K^-p\rightarrow\pi^0\Sigma^0$
reaction with the chiral-quark model and discussed characteristics
of the well established $\Lambda$ resonances. Liu et
al.~\cite{Xie_eta} analyzed the $K^-p\rightarrow\eta\Lambda$
reaction~\cite{CBdata0} with an effective Lagrangian approach and
implied a D03 resonance with mass about 1670 MeV but much smaller
width compared with the well established $\Lambda(1690){3\over
2}^-$. So there are still some ambiguities of the $\Lambda$ resonant
structures needing to be clarified.

Recently, the most precise data on the differential cross sections
for the $K^-p\to\pi^0\Sigma^0$ reaction have been provided by the
Crystal Ball experiment at AGS/BNL~\cite{CBdata1,CBdata2}. The
$\Sigma^0$ polarization data were presented for the first time.
However, with different data selection cuts and reconstructions, two
groups in the same collaboration, {\sl i.e.}, VA
group~\cite{CBdata2} and UCLA group~\cite{CBdata1}, got inconsistent
results for the $\Sigma^0$ polarizations. Previous multi-channel
analysis-\cite{Zhang:2013sva,Kamano:2015hxa,xhZhong} of the
$\overline{K}N$ reactions failed to reproduce either set of the
polarization data.

In our recent work~\cite{Shi:2014vha}, we concentrate on the most
precise data by the Crystal Ball collaboration on the pure isospin
scalar channel of $\overline{K}N$ reaction to see what are the
$\Lambda$ resonances the data demand and how the two groups'
distinct polarization data~\cite{CBdata1,CBdata2} influence the
spectroscopy of $\Lambda$ resonances. Consistent differential cross
sections of earlier work by Armenteros et al.~\cite{LowEnergyData}
at lower energies are also used. It is found that the 4-star
$\Lambda(1670){1\over 2}^-$ and 3-star $\Lambda(1600){1\over 2}^+$
resonances listed in PDG~\cite{pdg} are definitely needed no matter
which set of CB data is used. In addition, there is strong evidence
for the existence of a new $\Lambda({3\over 2}^+)$ resonance around
1680 MeV no matter which set of data is used. It gives large
contribution to this reaction, replacing the contribution from the
4-star $\Lambda(1690){3\over 2}^-$ resonance included by previous
fits to this reaction.

Replacing the PDG $\Lambda(1690){3\over 2}^-$ resonance by a new
$\Lambda(1680){3\over 2}^+$ resonance has important implications on
hyperon spectroscopy and its underlying dynamics. While the
classical qqq constituent quark model~\cite{capstick} predicts the
lowest $\Lambda({3\over 2}^+)$ resonance to be around 1900 MeV in
consistent with the $\Lambda(1890){3\over 2}^+$ listed in PDG, the
penta-quark dynamics~\cite{pentq1} predicts to be below 1700 MeV in
consistent with $\Lambda(1680){3\over 2}^+$ claimed in this work.

A recent analysis~\cite{Xie_eta} of CB data on the
$K^-p\to\eta\Lambda$ reaction requires a $\Lambda({3\over 2}^-)$
resonance with mass about 1670 MeV and width about 1.5 MeV instead
of the well established $\Lambda(1690){3\over 2}^-$ resonance with
width around 60 MeV. Together with $N^*(1520){3\over 2}^-$,
$\Sigma(1542){3\over 2}^-$ suggested in Ref.~\cite{pzgao} and either
$\Xi(1620)$ or $\Xi(1690)$, they fit in a nice $3/2^-$ baryon nonet
with large penta-quark configuration, {\sl i.e.}, $N^*(1520)$ as
$|[ud]\{uq\}\bar q>$ state, $\Lambda(1520)$ as $|[ud]\{sq\}\bar q>$
state, $\Lambda(1670)$ as $|[ud]\{ss\}\bar s>$ state, and
$\Xi(16xx)$ as $|[ud]\{ss\}\bar q>$ state. Here $\{q_1q_2\}$ means a
diquark with configuration of flavor representation ${\bf 6}$, spin
1 and color $\bar 3$. The $\Lambda(1670)$ as $|[ud]\{ss\}\bar s>$
state gives a natural explanation for its dominant $\eta\Lambda$
decay mode with a very narrow width due to its very small phase
space meanwhile a D-wave decay~\cite{ZouHyp}.

Recent analyses~\cite{Kamano:2015hxa,Fernandez-Ramirez:2015tfa} also
support possible existence of the $\Lambda(1680){3\over 2}^+$, but
with a narrower width.

\section{SUMMARY and prospects}\label{summary}

Taking into account new data from Crystal Ball
(CB)~\cite{CBdata0,CBdata1,CBdata2}, LEPS~\cite{LEPS} and
CLAS~\cite{CLAS}, new analyses show strong evidences for the lowest
$\Sigma^*(1/2^-)$ to be in the range of $1380\sim 1500$ MeV, the
lowest $\Sigma^*(3/2^-)$ to be around 1550 MeV and the lowest
$\Lambda^*(3/2^+)$ to be around 1680 MeV. There is also evidence for
a very narrow $\Lambda^*(3/2^-)$ around 1670 MeV decaying to
$\Lambda\eta$. All these new hyperon resonances fit in the expected
pattern of unquenched quark models very well. It is very important
to pin down the existence of these new resonances.

Various processes could be used to study these hyperon resonances.
The neutrino induced hyperon production processes $\bar{\nu}_{e/\mu}
+ p \to e^+/\mu^+ + \pi + \Lambda/\Sigma$ may provide a unique clean
place for studying low energy $\pi\Lambda/\Sigma$ interaction and
hyperon resonances below $KN$ threshold~\cite{Wu:2013kla}. With
plenty production of $\Lambda_c$ at BESIII, JPARC, BelleII,
$\Lambda_c^+\to\pi^+\pi^0\Lambda$ could also be used to study
$\Sigma^*$. The $K^-$, $K_L$ beam experiments at JPARC and Jlab
could provide an elegant new source for $\Lambda^*$, $\Sigma^*$ and
$\Xi^*$* hyperon spectroscopy.  $K_Lp\to\Lambda\pi^+$,
$\Sigma^0\pi^+$, $\Sigma^+\pi^0$, $\Sigma^{*0}\pi^+$,
$\Sigma^{*+}\pi^0$ could pin down the $\Sigma^*(1540) 3/2^-$; $K_Lp
\to\Sigma^0\pi^0\pi^+$, $\Lambda\pi^0\pi^+$ could shed light on the
$\Sigma*(1380\sim 1500)1/2^-$, $\Sigma^*(1540)3/2^-$,
$\Lambda^*(1680)3/2^+$; $K_Lp\to\Sigma^0\eta\pi^+$,
$\Lambda\eta\pi^+$ may check $\Sigma^*(1380\sim 1500)1/2^-$,
$\Sigma^*(1540)3/2^-$, $\Lambda^*(1670)3/2^-$. We believe the
proposed $K_L$ beam experiments at JLAB could settle down the
spectrum of the low excited hyperon states which provide
complimentary information to the study of penta-quark states with
hidden charm~\cite{Wu:2010jy,Aaij:2015tga} and play a crucial role
for understanding the hadron dynamics and hadron structure.

\begin{acknowledgments}
I thank S.~Dulat, Puze~Gao, Jun~Shi, J.~J.~Wu, J.~J.~Xie for
collaboration works reviewed here. This work is supported by the
National Natural Science Foundation of China under Grant 11261130311
(CRC110 by DFG and NSFC).
\end{acknowledgments}

\newcommand{\etal}{{\em et al.}}

\end{document}